\documentclass[11pt,a4paper,svgnames]{article}
\usepackage[left=1.28cm,right=1.28cm,top=1.27cm,bottom=1.28cm]{geometry}
\usepackage{graphicx}
\usepackage[font=footnotesize,labelfont=bf]{caption}
\usepackage{amsfonts}
\usepackage{amsmath}
\usepackage{multicol}
\usepackage[english]{babel}
\usepackage[square,compress]{natbib}
\usepackage[scaled=0.92]{helvet}
\usepackage{subcaption}
\DeclareCaptionSubType[Alph]{figure}
\DeclareCaptionLabelFormat{bold}{\bfseries #1#2}

\usepackage{tikz}
\usetikzlibrary{calc,positioning,fit}

\setcitestyle{numbers}
\let\OLDthebibliography\thebibliography
\renewcommand\thebibliography[1]{
  \OLDthebibliography{#1}  \setlength{\parskip}{0pt}
  \setlength{\itemsep}{0pt plus 0.3ex}
}

\newcommand{\VM}{\mathrm{VM}}

\newcommand\blfootnote[1]{%
  \begingroup
  \renewcommand\thefootnote{}\footnote{#1}%
  \addtocounter{footnote}{-1}%
  \endgroup
}

\begin{document}
\begin{Large} \begin{center}
    {\textbf{ Attraction to hierarchical feature memory explains orientation bias}} 
\end{center}
\end{Large}
\vspace{-1em}
\begin{large}
\begin{center}
{Kira M. D\"usterwald\textsuperscript{1,*}, Peter Vincent\textsuperscript{2}, Ana Kapros\textsuperscript{1}, Athena Akrami\textsuperscript{2} and Maneesh Sahani\textsuperscript{1}
\\
\vspace{0.1em}
\small \textit{\textsuperscript{1} Gatsby Computational Neuroscience Unit and \small \textsuperscript{2} Sainsbury Wellcome Centre, University College London}}
\end{center}
\end{large}
\vspace{-1.5ex}
\begin{multicols}{2}

\noindent \normalsize 
\textbf{Summary:} When recalling the orientation of recent stimuli, observers are systematically biased away from the cardinal axes. The prevailing explanation is that this “anti-cardinal bias” arises because cardinal orientations are encoded with greater neural resources and therefore less noise, consistent with efficient coding of environmentally common features. Under this account, the bias should occur independently for each stimulus; any serial attraction towards previously seen orientations should be weak and independent of absolute orientation---an interpretation supported by earlier behavioural reports.
By contrast, we find that serial effects in orientation recall are unexpectedly strong and reflect a hitherto unrecognised hierarchical interaction.  Sensory systems are organised in hierarchies, with higher-level areas representing complex combinations of lower-level features.  Neurons connect recurrently across the hierarchy, and we hypothesised that memories of simple features may be shaped by these cross-level interactions. Indeed, we found that serial dependence in visual orientation recall is dominated not by attraction to the previous orientation itself, but to a compound feature representing both a tilt and its mirror reflection about the cardinal axes.
Consistent with greater spatial abstraction at higher levels, the influence of this compound feature persisted across the visual midline, whereas orientation-specific effects weakened between hemifields. We developed a quantitative model based on the optimal combination of a noisy representation of the current stimulus with a hierarchical memory trace incorporating these compound features.  Remarkably, although the coding fidelity was uniform across angles, this model accurately reproduced not only the serial dependence, but also the anti-cardinal bias.
Our results suggest that memory traces are not isolated to single features, but reflect compound reactivation of hierarchical representations.  
Furthermore, they prompt a revision of the standard account of orientation biases, revealing that inference across hierarchical representations can generate systematic perceptual distortions previously ascribed to coding constraints.

\begin{figure}[!tbp]
\centering
\noindent%
%
\begin{tikzpicture}[x={(0,\linewidth)}, y={(0,0.5\textheight)},
    panel/.style={inner sep=0pt, minimum height=0.13\textheight},
    panel label/.style={inner sep=0pt, below right, font={\small\bfseries\sffamily}},
    panel title/.style={below, font={\tiny\sffamily}},
    panel xlabel/.style={above, font={\tiny\sffamily}},
    panel ylabel/.style={rotate=90, below, font={\tiny\sffamily}},
    panel rlabel/.style={rotate=90, above, font={\tiny\sffamily}},
  ]
  \def\plabel#1{\gdef\thispanel{#1}\node[panel label] at (#1.north west) {#1}}
  \newcommand\ptitle[2][0,0]{\node[panel title]  at ($(\thispanel.north)+(#1)$) {#2}}
  \newcommand\xlabel [2][0,0]{\node[panel xlabel] at ($(\thispanel.south)+(#1)$) {#2}}
  \newcommand\ylabel [2][0,0]{\node[panel ylabel] at ($(\thispanel.west) +(#1)$) {#2}}
  \newcommand\rlabel[2][0,0]{\node[panel rlabel]at ($(\thispanel.east) +(#1)$) {#2}}
  \renewcommand{\-}{\text{--}}
  \newcommand\prev{_{n\-1}}

  \node[panel, below right] (A) at (0, 1) {\includegraphics[width=0.27\textwidth]{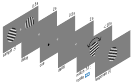}};
  \plabel A;

  \node[panel, below right, xshift=2pt] (B) at (A.north east) {\includegraphics[height=0.11\textheight]{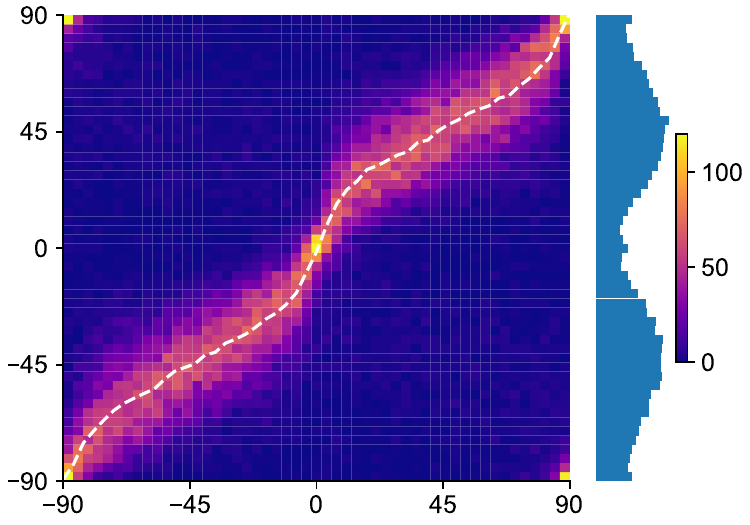}};
  \plabel B;
  \ptitle[-1em,0] {joint histogram};
  \xlabel[-1em,0] {sample $S_n$};
  \ylabel[-6pt,0] {response $R_n$};
  \rlabel[ 5pt,0] {count};
  \draw[<-,white] (B.center) ++(4pt,18pt) -- ++(-5pt,5pt) node [pos=1,panel title, inner sep=0pt,above left] {bias $\mathbb E[R_n\-S_n]$};

  \node[panel, below right, xshift=6pt] (C) at (B.north east) {\includegraphics[width=0.22\textwidth]{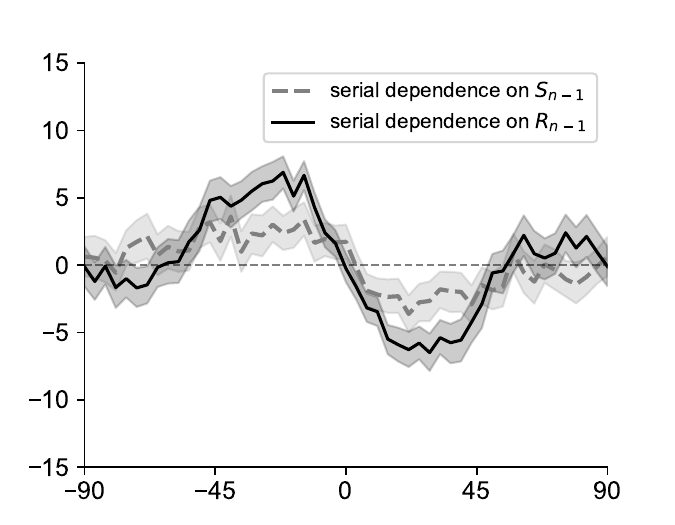}};
  \plabel C;
  \xlabel[0pt,-2pt] {relative orientation $S_n\-\{S\prev,R\prev\}$};
  \ylabel[-3pt,0] {bias $\mathbb E[R_n\-S_n]$};

  \node[panel, below right, xshift=-10pt, minimum width=10em] (G) at (C.north east)
       {\includegraphics[width=0.215\textwidth,clip]{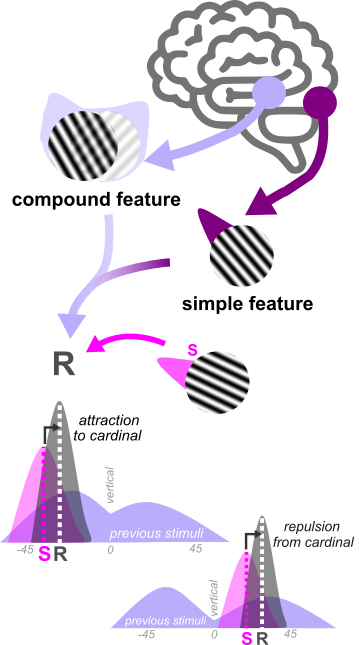}};
  \plabel G;
  \node[fill=white,xshift=-35pt,yshift=-13pt,inner sep=0pt] at (\thispanel) {$R$};

  \node[panel, below right] (D) at (A.south west) {\raisebox{1ex}{\includegraphics[height=0.13\textheight,trim=0 0 71 0,clip]{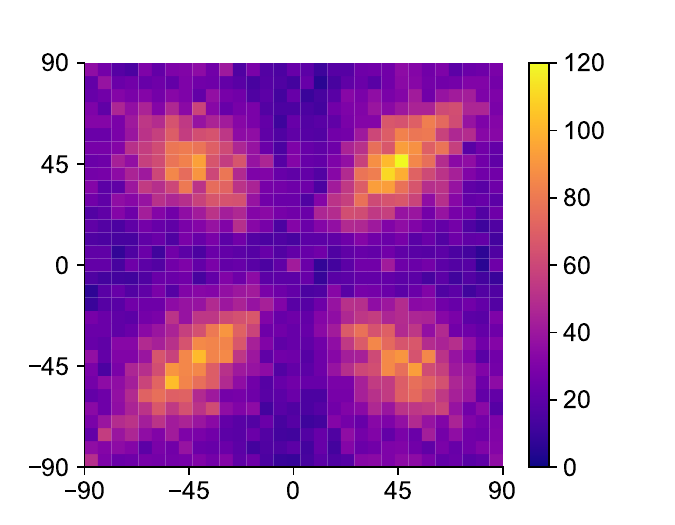}}};
  \begin{scope}[shift=(D),x=(D.east),y=(D.north),every node/.style={rotate=45,white,font={\tiny\bfseries\sffamily}}]
    \node at (0.1 ,0.5) {r=--0.11};
    \node at (0.75,0.5) {r=0.17};
    \node at (0.1 ,-0.2) {r=0.2};
    \node at (0.75,-0.2) {r=--0.1};
  \end{scope}
  \plabel D;
  \ptitle[0.5em,0] {joint histogram $(R\prev,R_n)$};
  \xlabel[0.7em,0] {previous response $R\prev$};
  \ylabel[-2pt,2pt] {response $R_n$};

  \node[panel, below right, xshift=15pt] (E) at (D.north east){\raisebox{1ex}{\includegraphics[height=0.13\textheight,trim=0 0 20 0,clip]{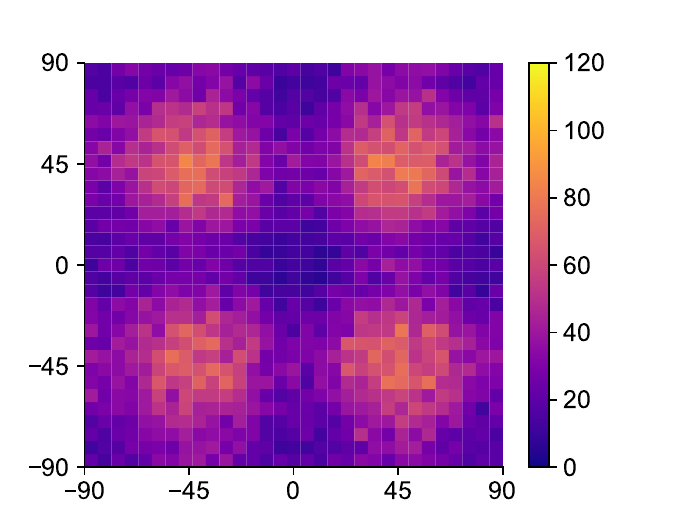}}};
  \plabel E;
  \begin{scope}[shift=(\thispanel),x=(\thispanel.east),y=(\thispanel.north),every node/.style={rotate=45,white,font={\tiny\bfseries\sffamily}}]
    \node at (-.1 ,0.5) {r=0.01};
    \node at (0.45,0.5) {r=0.01};
    \node at (-.1 ,-0.2) {r=0.0};
    \node at (0.45,-0.2) {r=0.0};
  \end{scope}
  \ptitle {resampled histogram $(R_{n'},R_n)$\hspace*{2em}};
  \xlabel {resampled $R_{n'}$};
  \ylabel[-2pt,2pt] {response $R_n$};
  \rlabel {count};

  \node[panel, below right, xshift=6pt, minimum width=10em] (F) at (E.north east) {\raisebox{1ex}{\includegraphics[height=0.13\textheight,trim=0 0 20 0,clip]{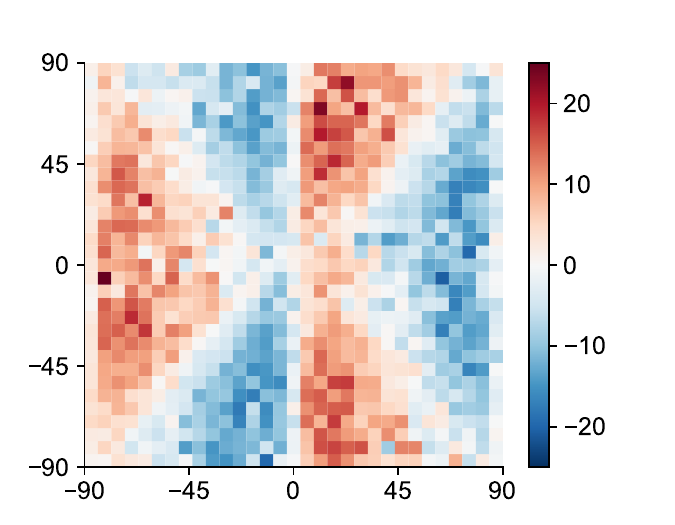}}};
  \plabel F;
  \ptitle {$\mathbb{E}[R_n\-S_n|S_n,R\prev]$ in data};
  \xlabel[-1pt,0]{sample $S_n$};
  \ylabel[-4pt,3pt] {previous response $R\prev$};
  \rlabel {degrees};

  \node[panel, below right] (H) at (D.south west) {\raisebox{1ex}{\includegraphics[height=0.13\textheight,trim=0 0 71 0,clip]{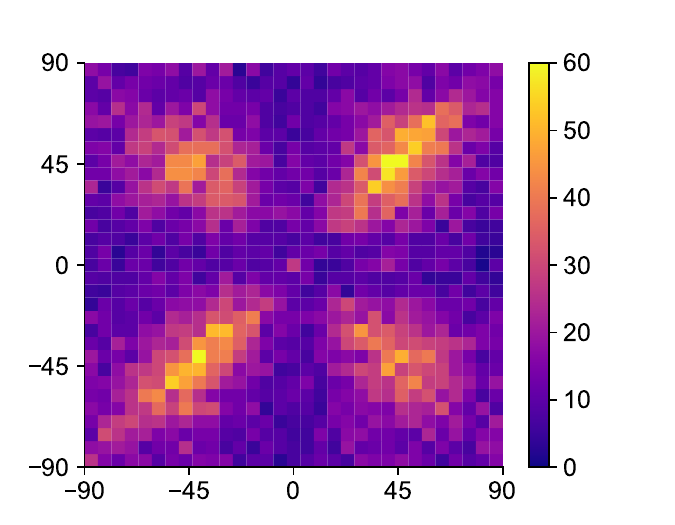}}};
  \plabel H;
  \begin{scope}[shift=(\thispanel),x=(\thispanel.east),y=(\thispanel.north),every node/.style={rotate=45,white,font={\tiny\bfseries\sffamily}}]
    \node at (0.1 ,0.5) {r=--0.1};
    \node at (0.75,0.5) {r=0.19};
    \node at (0.1 ,-0.2) {r=0.24};
    \node at (0.75,-0.2) {r=--0.11};
  \end{scope}
  \ptitle[4pt,-2pt] {$S\prev$ on same side as $S_n$};
  \xlabel[0.7em,0] {previous response $R\prev$};
  \ylabel[-2pt,2pt] {response $R_n$};

  \node[panel, below right,xshift=15pt] (I) at (H.north east) {\raisebox{1ex}{\includegraphics[height=0.13\textheight,trim=0 0 20 0,clip]{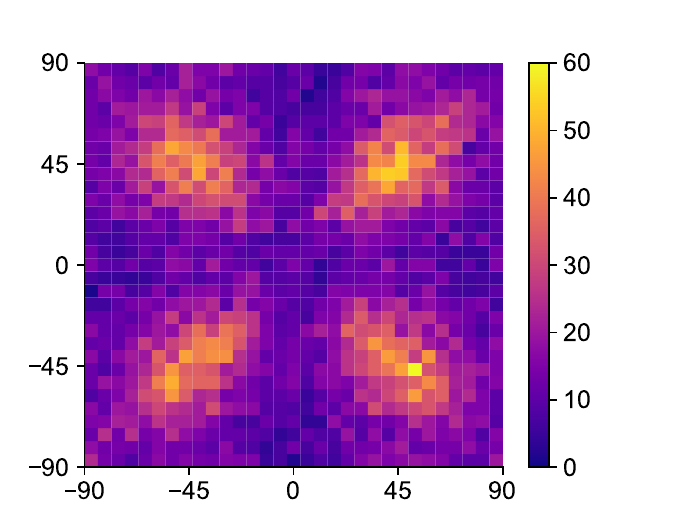}}};
  \plabel I;
  \begin{scope}[shift=(\thispanel),x=(\thispanel.east),y=(\thispanel.north),every node/.style={rotate=45,white,font={\tiny\bfseries\sffamily}}]
    \node at (-.1 ,0.5) {r=--0.12};
    \node at (0.45,0.5) {r=0.15};
    \node at (-.1 ,-0.2) {r=0.17};
    \node at (0.45,-0.2) {r=--0.08};
  \end{scope}
  \ptitle[-4pt,-2pt] {$S\prev$ on opposite side to $S_n$};
  \xlabel {previous response $R\prev$};
  \ylabel[-2pt,2pt] {response $R_n$};
  \rlabel {count};

  \node[panel, below right,xshift=6pt,minimum width=10em] (J) at (I.north east) {\raisebox{1ex}{\includegraphics[height=0.13\textheight,trim=0 0 20 0,clip]{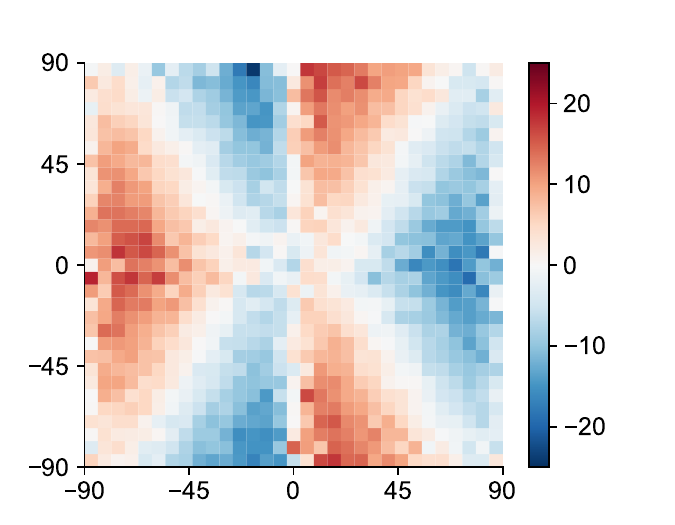}}};
  \plabel J;
  \ptitle {$\mathbb{E}[R_n\-S_n|S_n,R\prev]$ in hier.\ model};
  \xlabel[-1pt,0]{sample $S_n$};
  \ylabel[-4pt,3pt] {previous response $R\prev$};
  \rlabel {degrees};

  \node[panel, below right, xshift=13pt] (K) at (J.north east) {\raisebox{1ex}{\includegraphics[height=0.13\textheight,trim=0 0 20 0,clip]{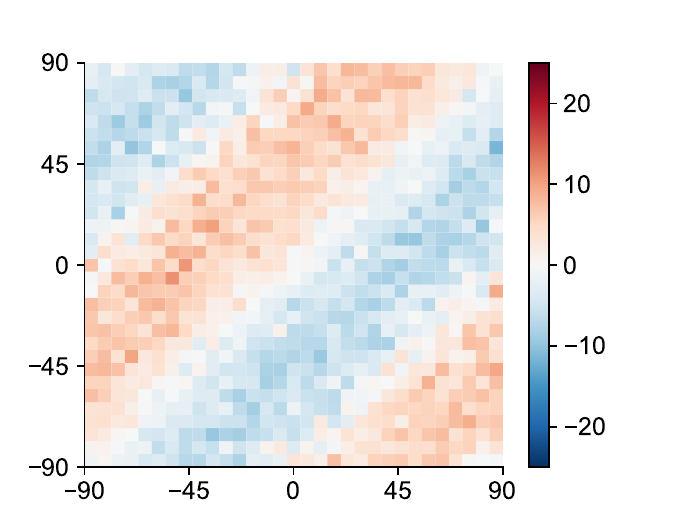}}};
  \plabel K;
  \ptitle {$\mathbb{E}[R_n\-S_n|S_n,R\prev]$ in 1-feat.\ model};
  \xlabel[-1pt,0] {sample $S_n$};
  \ylabel[-4pt,3pt] {previous response $R\prev$};
  \rlabel {degrees};

  \node[panel, below right,xshift=5pt] (L) at (H.south west) {\includegraphics[width=0.22\textwidth]{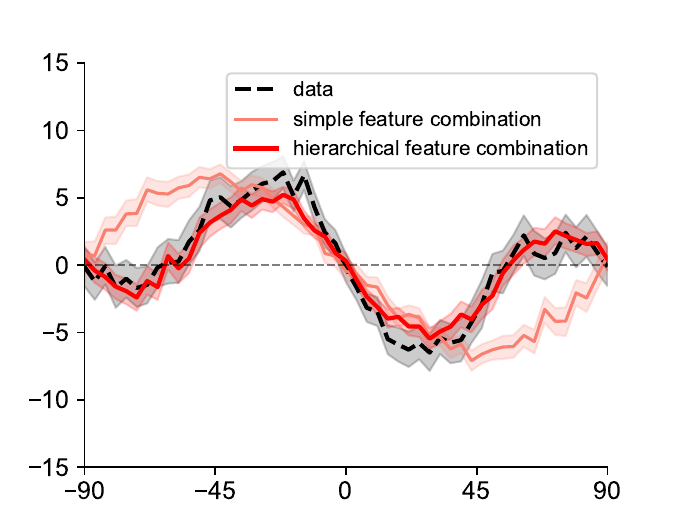}};
  \plabel L;
  \xlabel[0pt,-2pt] {relative orientation $S_n\-\{S\prev,R\prev\}$};
  \ylabel[-3pt,0] {bias $\mathbb E[R_n\-S_n]$};

  \node[panel, below right, xshift=5pt] (M) at (L.north east) {\includegraphics[width=0.22\textwidth]{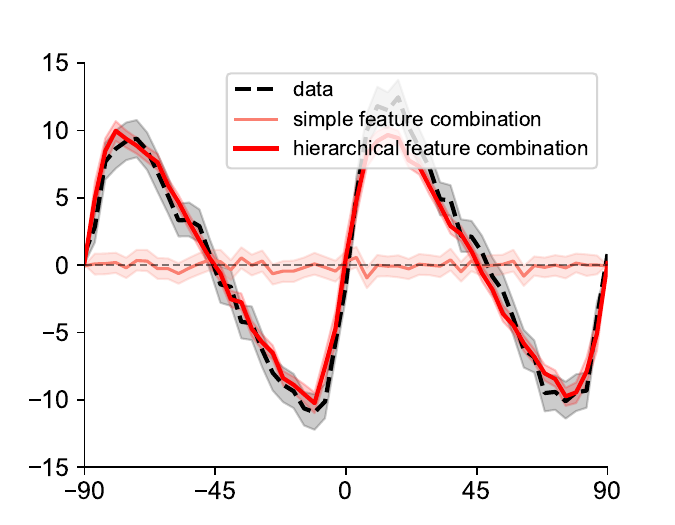}};
  \plabel M;
  \xlabel {sample $S_n$};
  \ylabel[-3pt,0] {bias $\mathbb E[R_n\-S_n]$};

  \node[panel, below right, xshift=2pt] (N) at (M.north east)
  {\raisebox{1ex}{\includegraphics[height=0.13\textheight,trim=0 0 20 0,clip]{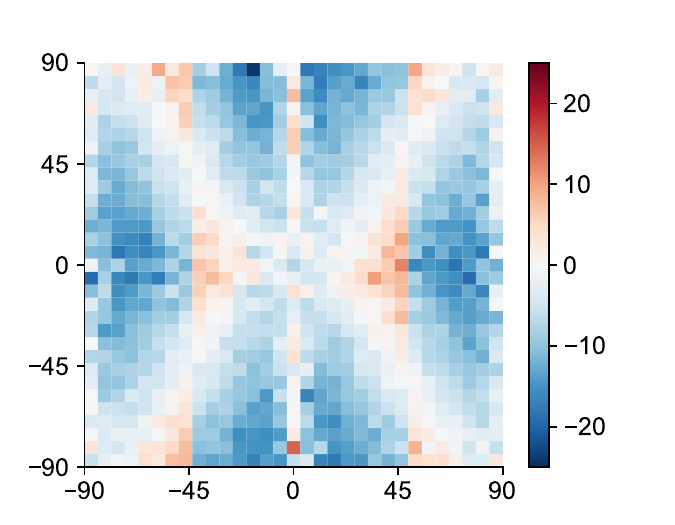}}};
  \plabel N;
  \begin{scope}[shift={($(\thispanel.center)+(1pt,2pt)$)},
                x=($0.8*(\thispanel.east)$),y=($0.77*(\thispanel.north)$)]
    \node[font={\tiny\sffamily},text width=3em,align=center, gray] at (-0.5, 0.48) {attracted to\\ reflection};

    \fill[gray,opacity=0.25] (-0.12,0) rectangle (-0.95,-1.03);
    \node[font={\tiny\sffamily},text width=3em,align=center] at (-0.5, -0.5) {attracted to prev.\ resp.};
    
    \node[font={\tiny\sffamily},text width=3em,align=center, gray] at (0.35, -0.45) {attracted to\\ reflection};

    \fill[gray,opacity=0.25] (-0.12,0) rectangle (0.73,1.03);
    \node[font={\tiny\sffamily},text width=3em,align=center] at (0.35, 0.5) {attracted to prev.\ resp.};

  \end{scope}
  
  \ptitle {anti-cardinal bias in hier.\ model};
  \xlabel[-1pt,0]{sample $S_n$};
  \ylabel[-4pt,3pt] {previous response $R\prev$};

  \fill[white] (\thispanel.south east) rectangle ($(\thispanel.north east)+(-15pt,0)$);
  \draw[->] (\thispanel.east) ++(-10pt,5pt) -- ++(0,1cm) node[panel ylabel,midway] {pro-cardinal};
  \draw[->] (\thispanel.east) ++(-10pt,-3pt) -- ++(0,-1cm) node[panel ylabel,midway] {anti-cardinal};

  \node[panel, below right, xshift=13pt] (O) at (N.north east) {\includegraphics[height=0.13\textheight]{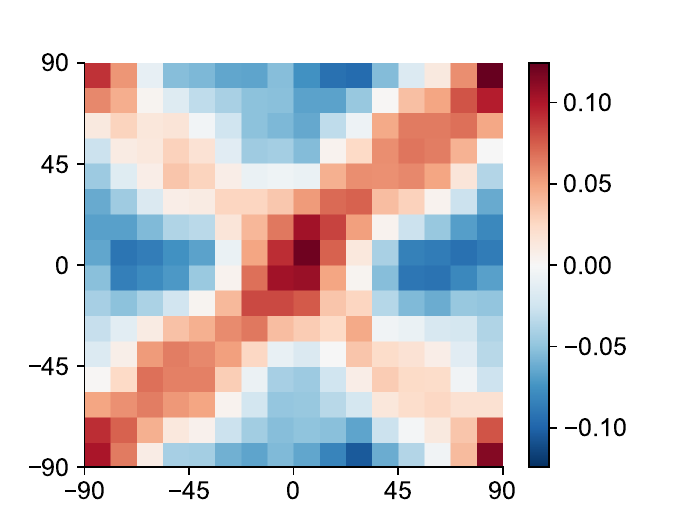}};
  \plabel O;
  \ptitle[-0.7em,0] {ERP similarity};
  \xlabel[-5pt,-5pt] {sample $S_n$};
  \ylabel[-5pt,0pt] {reference $\{S_{n'}\}$};
  \rlabel {norm.\ mahanalobis distance};
    
\end{tikzpicture}
\begin{subcaptiongroup}  
\phantomcaption\label{fsyn:expt}%
\phantomcaption\label{fsyn:SnRnhist}%
\phantomcaption\label{fsyn:serialbias}%
\phantomcaption\label{fsyn:Rn-1Rn:data}%
\phantomcaption\label{fsyn:Rn-1Rn:shuffle}%
\phantomcaption\label{fsyn:jointbias:data}%
\phantomcaption\label{fsyn:schematic}%
\phantomcaption\label{fsyn:Rn-1Rn:ipsi}%
\phantomcaption\label{fsyn:Rn-1Rn:contra}%
\phantomcaption\label{fsyn:jointbias:hier}%
\phantomcaption\label{fsyn:jointbias:feat}%
\phantomcaption\label{fsyn:serialbias:models}%
\phantomcaption\label{fsyn:bias:models}%
\phantomcaption\label{fsyn:bias:proanti}%
\phantomcaption\label{fsyn:erpdecoding}%
\end{subcaptiongroup}
\captionsetup{subrefformat=bold}
\caption[]{Attraction to hierarchical feature memory explains orientation bias.  
  (\subref{fsyn:expt})~Task: $86$ participants (after exclusions) performed 400 trials each online. Trials began with two sample grating stimuli at orientations drawn independently and uniformly around the circle.  After a delay, one grating location was cued; after a further delay, a randomly oriented manipulandum grating appeared and participants used arrow keys to rotate this to match the recalled orientation at the cued location. Data 
are pooled across all participants.
(\subref{fsyn:SnRnhist})~Cued samples $S_n$ and behavioural responses $R_n$ generate a joint histogram with a bimodal $R_n$ marginal.  Responses are biased towards oblique angles (dashed white line).  Here, and throughout, 0 corresponds to vertical.
(\subref{fsyn:serialbias})~The error on trial $n$ as a function of the orientation difference between current sample ($S_n$) and previous sample ($S_{n-1}$, grey) or response ($R_{n-1}$, black).  Negative abscissa values indicate that the previous value was more clockwise than $S_n$; positive errors mean that $R_n$ was on average more clockwise than $S_n$.  Shadings are bootstrapped 95\% confidence intervals.
(\subref{fsyn:Rn-1Rn:data})~Joint histogram of responses on trial $n$ and responses on the previous trial $n-1$; shared colour scale with \subref{fsyn:Rn-1Rn:shuffle}. Circular correlation coefficients for each quadrant are shown in white (also in \subref{fsyn:Rn-1Rn:shuffle},\subref{fsyn:Rn-1Rn:ipsi}--\subref{fsyn:Rn-1Rn:contra}). 
(\subref{fsyn:Rn-1Rn:shuffle})~Joint histogram of responses on trial $n$ and permuted responses (null distribution).
(\subref{fsyn:jointbias:data})~Average bias $\mathbb{E}[R_n-S_n]$ conditioned on $S_n$ and $R_{n-1}$.
(\subref{fsyn:schematic})~Schematic of hierarchical compound feature neural representation.  Remembered stimuli, conveyed by previous responses $R_{n-1}, R_{n-2},\dots$, leave implicit hierarchical memory traces combining a compound feature representing the remembered orientation and its cardinal-axis reflection (light purple) as well as a primary orientation-only signal (dark purple).
When subsequent stimuli are processed, existing hierarchical feature traces are reactivated, and combined with a simple belief about the current sample $S$ (pink). The modelled response $R$ is the mean of the combined beliefs.  Depending on the direction of the closest mode in the memory trace from $S$, this response may be biased towards or away from the cardinal axes.
%
(\subref{fsyn:Rn-1Rn:ipsi})~Joint histogram of $R_n$ and $R_{n-1}$ for trials in which $S_n$ occurred on the same side of the screen as $S_{n-1}$ (i.e. both left or both right), shares colour scale with \subref{fsyn:Rn-1Rn:contra}.
(\subref{fsyn:Rn-1Rn:contra})~As for \subref{fsyn:Rn-1Rn:ipsi}, but for the trials in which $S_n$ was on a different side of the screen to $S_{n-1}$ (left/right or right/left).
(\subref{fsyn:jointbias:hier})~As in \subref{fsyn:jointbias:data}: based on samples from hierarchical compound feature model.
(\subref{fsyn:jointbias:feat})~As in \subref{fsyn:jointbias:data}: based on samples from simple feature model.
(\subref{fsyn:serialbias:models})~Hierarchical feature (red) and simple feature (peach) model response errors as a function of $S_n-R_{n-1}$.  Data curve (black) from \subref{fsyn:serialbias}. Shadings are bootstrapped 95\% confidence intervals. The simple feature model fails to capture apparent repulsion at large differences.
(\subref{fsyn:bias:models})~Average response bias across stimuli $S_n$ in different models (colours, shadings as in \subref{fsyn:serialbias:models}), and compared to data from \subref{fsyn:SnRnhist}.
(\subref{fsyn:bias:proanti})~Pattern of hierarchical model bias as in \subref{fsyn:jointbias:hier}, but coloured by pro- or anti-cardinal bias direction.  Anti-cardinal effects predominate for all values of $S_n$. Annotations indicate which compound feature mode is closer to $S_n$ and so dominates attraction.
(\subref{fsyn:erpdecoding})~Re-analysis of EEG data from Wolff et al., 2020 \citep{Wolff2020}: Average ERP similarity (mean-centred Mahalanobis distances) between held-out cued sample orientations $S_n$ and sets $\{S_{n'}\}$ of template responses collected within orientation bins. 
} 
\label{fsyn}
\end{figure}
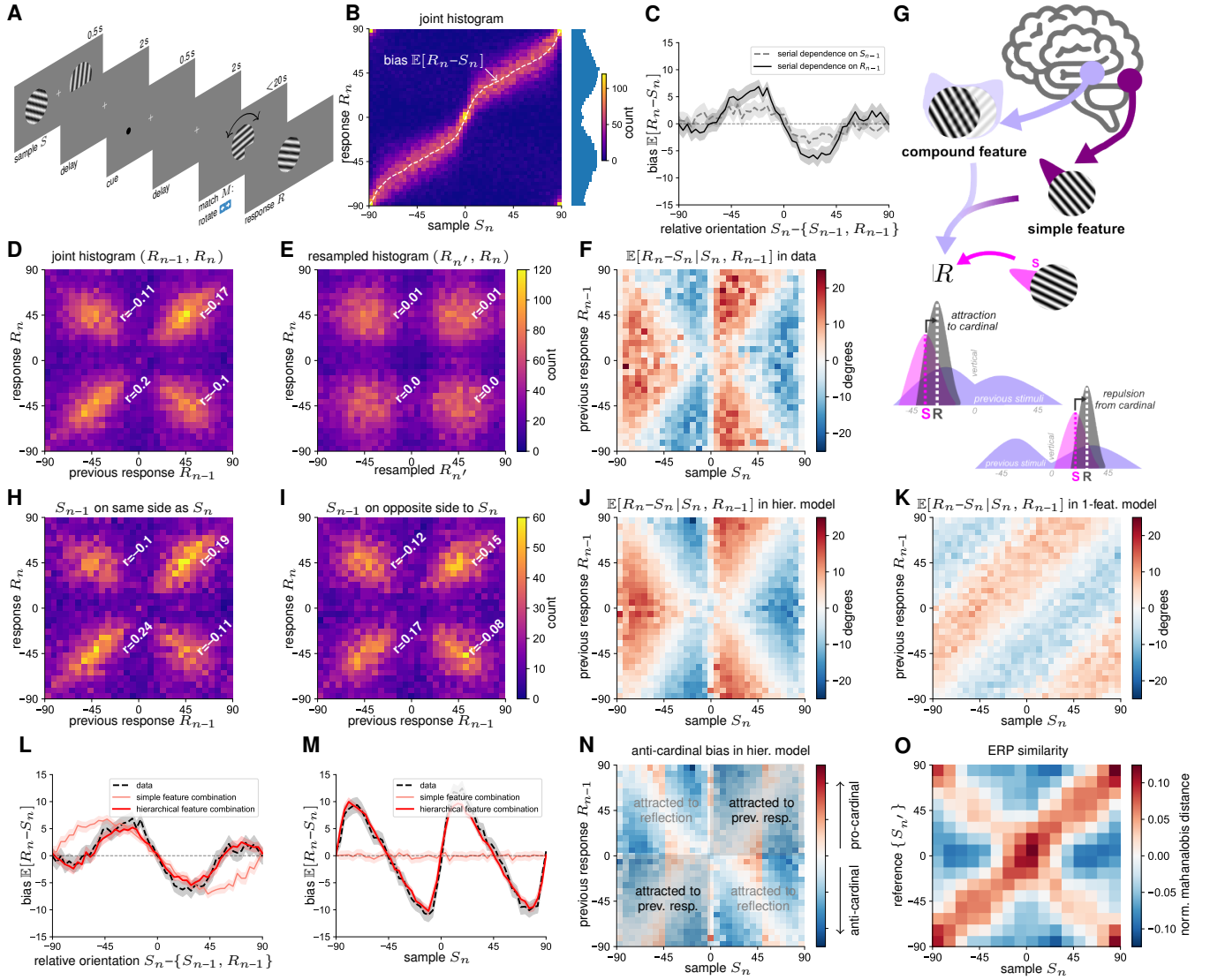

\vskip 1mm
\noindent\textbf{Additional detail:} $86$ online participants reproduced the orientation of briefly presented, randomly oriented gratings by rotating a match grating (Fig.~\ref{fsyn:expt}). Consistent with previous work, participants exhibited 
both a response bias towards oblique angles (\ref{fsyn:SnRnhist}) \citep{noel2021, Wei2015, Wolff2020} and mostly
attractive serial-dependence (\ref{fsyn:serialbias}) \citep{Fischer2014}.
Behavioural responses $R_n$ were more strongly biased towards the orientation reported on the previous trial ($R_{n-1}$) than to the cued sample ($S_{n-1}$), and a GAM analysis (not shown) confirmed that $S_{n-1}$  captured no further variance beyond that predicted by $R_{n-1}$.
We therefore quantified serial effects in terms of $R_{n-1}$.
%

%
Examining the influence of $R_{n-1}$ on $R_n$ directly (rather than as bias from $S_n$) revealed a remarkable and hitherto unreported effect.
When $R_{n-1}$ and $R_n$ fell on the same side of vertical (congruent), we found a significant positive circular correlation, indicating attraction (\ref{fsyn:Rn-1Rn:data}; $p<0.001$, permutation test on shuffled null, \ref{fsyn:Rn-1Rn:shuffle}).  However, when $R_{n-1}$ and $R_n$ fell on opposite sides (incongruent) there was a significant negative circular correlation ($p<0.001$), revealing attraction to the \emph{reflection} of $R_{n-1}$.
Indeed, bias as a function of $S_n$ and $R_{n-1}$ was similar for reflected values of $R_{n-1}$ (\ref{fsyn:jointbias:data}).

We hypothesised that this result may arise from the serial influence of a remembered hierarchical representation that included both low-level orientation \emph{and} an implicitly activated compound feature  jointly representing a tilt and its reflection across cardinal axes (\ref{fsyn:schematic}).  
As lower-level visual representations tend to be more spatially specific, we predicted that such a hierarchical influence would be sensitive to spatial configuration: indeed, when $S_{n-1}$ and $S_n$ were presented in opposite hemifields, congruent $R_n$ and $R_{n-1}$ were less strongly correlated, whereas the incongruent correlation---putatively arising from the compound factor---was maintained (\ref{fsyn:Rn-1Rn:ipsi}--\subref{fsyn:Rn-1Rn:contra}).

We modelled $R_n$ as the mean of a combined belief distribution: a von Mises belief about the current sample $S_n$ based on noisy internal estimate $X_n$, $p(S_n|X_n) = \VM(X_n, \kappa_X), X_n \sim \VM\!\left(S_n, \kappa_S\right)$, and von Mises mixtures representing the compound feature memory traces of $R_{n-1}$ and $R_{n-2}$.  Implicitly, the prior on encoding precision is uniform.  We fit the von Mises concentration parameters to the serial dependence bias (\ref{fsyn:serialbias}). 
The model reproduced the pattern of serial dependence (\ref{fsyn:jointbias:hier},\subref{fsyn:serialbias:models}), unlike a model that relied on simple feature attraction (\ref{fsyn:jointbias:feat},\subref{fsyn:serialbias:models}).
Remarkably, and despite uniform coding precision across angles, the same model reproduced the anti-cardinal response bias completely, without any further tuning (\ref{fsyn:bias:models}).  
We conclude that the long-studied anti-cardinal orientation bias arises in fact from the average anti-cardinal dominance of attraction to $R_{n-1}$ and its reflection (\ref{fsyn:bias:proanti}).

These results held for laboratory participants in our own study, and for two analogous publicly available datasets \citep{noel2021, Wolff2020}. Moreover, decoding re-analysis of EEG from \citep{Wolff2020} revealed that orientation-evoked potentials resembled both similar \emph{and} reflected orientations  (\ref{fsyn:erpdecoding}), providing preliminary evidence that working memory encodes compound tilt/reflection features at the neural level.
\bibliographystyle{unsrtabbrv}
\renewcommand{\bibsection}{}
\blfootnote{\vspace{-4ex}\begin{multicols}{2} \begin{footnotesize} \bibliography{ref}

@article{Fischer2014,
   author = {Jason Fischer and David Whitney},
   doi = {10.1038/NN.3689},
   issn = {15461726},
   issue = {5},
   journal = {Nat. Neurosci},
   pmid = {24686785},
   publisher = {Nature Publishing Group},
   volume = {17},
   year = {2014}
}

@article{Wei2015,
   author = {Xue Xin Wei and Alan A. Stocker},
   doi = {10.1038/nn.4105},
   issn = {15461726},
   issue = {10},
   journal = {Nat Neurosci},
   month = {10},
   pmid = {26343249},
   publisher = {Nature Publishing Group},
   
   volume = {18},
   year = {2015},
}

@article{noel2021,
    doi = {10.1371/journal.pbio.3001215},
    author = {J. P. Noel and others},
    journal = {PLoS Biol.},
    publisher = {Public Library of Science},
    year = {2021},
    month = {05},
    volume = {19},
    url = {https://doi.org/10.1371/journal.pbio.3001215},
    number = {5},

}

@article{Wolff2020,
  author = {Michael J. Wolff and others},
   doi = {10.1371/journal.pbio.3000625},
   issn = {15457885},
   issue = {3},
   journal = {PLoS Biol.},
   month = {3},
   pmid = {32119658},
   publisher = {Public Library of Science},
   volume = {18},
   year = {2020},
}
\vspace{-2ex} \begin{flushright} * {\textit{Correspondence:}  \texttt{kira.dusterwald.21@ucl.ac.uk} \\ Poster at Computational and Systems Neuroscience (COSYNE) conference, Lisbon, 2026} \end{flushright}\end{footnotesize} 
\end{multicols}}
\end{multicols} \vspace{-5em}

\end{document}